\documentclass[runningheads]{svmult}
\usepackage{graphicx}  

\begin{document}

\def\sin{{\rm sin}}
\def\cos{{\rm cos}}

\title*{Whispering Gallery Resonators and \ \protect\newline Tests of Lorentz Invariance}

\author{Peter Wolf\inst{1,2}
\and Michael E. Tobar\inst{3}
\and S\'{e}bastien Bize\inst{1}
\and Andr\'{e} Clairon\inst{1}
\and Andr\'{e} N. Luiten\inst{3}
\and Giorgio Santarelli\inst{1}}

\authorrunning{Peter Wolf et al.}

\institute{BNM-SYRTE, Observatoire de Paris, 61 avenue de l'Observatoire, 75014 Paris, France
\and Bureau International des Poids et Mesures, Pavillon de Breteuil, 92312 S\`evres Cedex, France
\and University of Western Australia, School of Physics, Nedlands 6907 WA, Australia}

\maketitle

\begin{abstract}
The frequencies of a cryogenic sapphire oscillator and a hydrogen maser are compared to set new constraints on a possible violation of Lorentz invariance. We give a detailed description of microwave resonators operating in Whispering Gallery modes and then apply it to derive explicit models for Lorentz violating effects in our experiment. Models are calculated in the theoretical framework of Robertson, Mansouri and Sexl (RMS) and in the standard model extension (SME) of Kostelecky and co-workers. We constrain the parameters of the RMS test theory to $1/2 - \beta_{MS} + \delta_{MS} = (1.2 \pm 2.2) \times 10^{-9}$ and $\beta_{MS} - \alpha_{MS} - 1 = (1.6 \pm 3.0) \times 10^{-7}$ which is of the same order as the best results from other experiments for the former and represents a 70 fold improvement for the latter. These results correspond to an improvement of our previously published limits [Wolf P. et al., Phys. Rev. Lett. {\bf 90}, 6, 060402, (2003)] by about a factor 2.
\end{abstract}

\section{Introduction}

The Einstein equivalence principle (EEP) is at the heart of special and general relativity \cite{Will} and a cornerstone of modern physics. One of the constituent elements of EEP is Local Lorentz invariance (LLI) which, loosely stated, postulates that the outcome of any local test experiment is independent of the velocity of the freely falling apparatus (the fundamental hypothesis of special relativity). The central importance of this postulate in modern physics has motivated tremendous work to experimentally test LLI \cite{Will}. Additionally, nearly all unification theories (in particular string theory) violate the EEP at some level \cite{KostoSam,Damour1} which further motivates experimental searches for such violations of the universality of free fall \cite{Damour2} and of Lorentz invariance \cite{Kosto1,Kosto2}.

Numerous test theories that allow the modeling and interpretation of experiments that test LLI have been developed. Kinematical frameworks \cite{Robertson,MaS} postulate a simple parametrisation of the Lorentz transformations with experiments setting limits on the deviation of those parameters from their special relativistic values. A more fundamental approach is offered by theories that parametrise the coupling between gravitational and non-gravitational fields (TH$\epsilon\mu$ \cite{LightLee,Will,Blanchet} or $\chi$g \cite{Ni} formalisms) which allow the comparison of experiments that test different aspects of the EEP. Formalisms based on string theory \cite{KostoSam,Damour1,Damour2} have the advantage of being well motivated by theories of physics that are at present the only candidates for a unification of gravity and the other fundamental forces of nature. Fairly recently a general Lorentz violating extension of the standard model of particle physics (Standard Model Extension, SME) has been developed \cite{Kosto1} whose Lagrangian includes all parametrised Lorentz violating terms that can be formed from known fields. Many of the theories mentioned above are included as special cases of the SME \cite{KM}.

We report here on experimental tests of LLI using a cryogenic
sapphire oscillator and a hydrogen maser. After a detailed description of microwave resonators operating in Whispering Gallery modes (Sect. 2) we explicitly calculate models for Lorentz violating effects in our experiment in the theoretical frameworks of \cite{MaS} (Sect. 3.1) and the SME (Sect. 3.2). In both cases the relative frequency
of the two clocks is modulated with, typically, sidereal
and semi-sidereal periods due to the movement of the lab with the
rotation of the Earth. The experimental results searching for those variations are presented and compared to previously published ones in Sect. 4. We set limits on parameters that describe
such Lorentz violating effects, improving our previous results \cite{Wolf} by a factor 2. Our limits are of the same order as the best results from other experiments for the Michelson-Morley type test \cite{Muller}, and correspond to a 70-fold improvement over the best other results \cite{Schiller} for the Kennedy-Thorndike type test.

\section{Whispering Gallery mode resonators}

\subsection{Basic principle}

\begin{figure}[htb]
\begin{center}
\includegraphics[width=4cm]{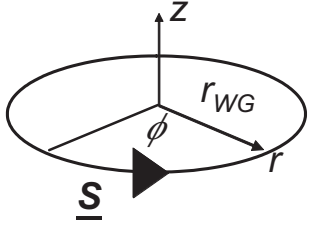}
\end{center}
\caption[]{Schematic of the propagation of a 'pure WG' mode in a cylindrical symmetric structure of radius $r_{WG}$.}
\label{fig:Mike1}
\end{figure}

Whispering Gallery (WG) modes in spherical and cylindrical dielectric and cavity resonators have been long studied, and have many applications ranging from novel filter designs to high-Q resonators for oscillator applications \cite{TobMan,Cros,Guillon,Krupka,Beyer,TobAns}. The boundary conditions ensure that the mode propagates mainly around the azimuth in the equatorial plane (in both directions in the case of a standing wave). A 'pure' WG mode, as shown in Fig. \ref{fig:Mike1} may be defined with only propagation in the azimuthal direction. Thus, for the 'pure' WG mode, the Poynting vector has only an $S_\phi$ component. The mode can be clasified as Transverse Electric (TE) or Transverse Magnetic (TM). TE modes have only $E_r$ and $H_z$ components, while TM modes have only $H_r$ and $E_z$ components.

For a dielectric WG mode resonator electromagnetic energy propagates just inside the dielectric-free space interface, due to total internal reflections, and if we assume the total internal reflection is perfect then we may assume 100 \% of the energy is within the dielectric. Boundary conditions dictate that only certain frequencies of resonance occur when an integral number of wavelengths fit within the resonator. This happens when

\begin{equation}
\nu_{wg}=\frac{m c}{2 \pi r_{wg}\sqrt{\epsilon}}=\frac{k_m}{t_c}
\label{fWG}
\end{equation}

where $m$ is an integer, which represents the number of wavelengths that fit along the circumference of the resonator, $c$ is the speed of light in vacuum, $\epsilon$ the relative permittivity of the dielectric, $k_m$ a constant, and $t_c$ the travel time (in vacuum) of a light signal around the circumference of the resonator.

\begin{figure}[htb]
\begin{center}
\includegraphics[width=10.5cm]{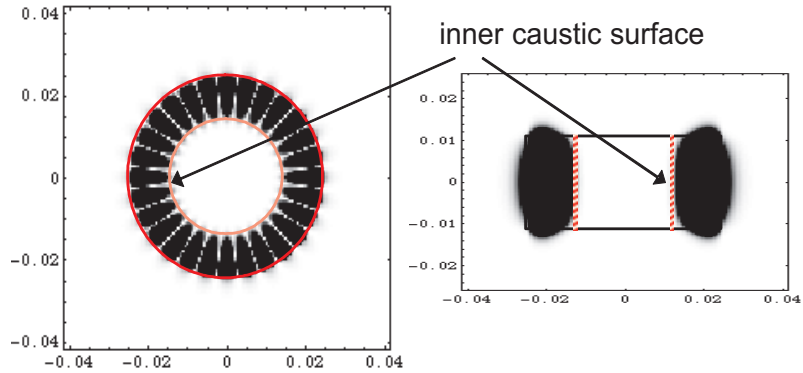}
\end{center}
\caption[]{Magnitude of the $H_z$ field calculated for the WGE$_{14,0,0}$ mode in a sapphire disk resonator using the separation of variables technique. The r-$\phi$ and r-z planes are represented. The inner caustic is shown, and the mode can be approximated as two guided waves propagating in opposite directions around the azimuth.}
\label{fig:Mike2}
\end{figure}

This relation is only true for 'pure' WG modes, which only occur as $m \to \infty$. At lower values of $m$ the path length is extended by reflections from an internal caustic surface (see Fig. \ref{fig:Mike2}), and in fact the wave propagates like a guided wave around the azimuth. In this case Maxwell's equations must be solved, and to calculate the mode frequencies accurately, the roots of the corresponding transcendental equation must be found \cite{TobMan}. For this case the modes have all electric and magnetic field components, and are classified by the dominant field components. Quasi-TE modes have dominant $E_r$ and $H_z$ field components, and are usually denoted as WGE or $H_z$ modes. Conversely, modes with dominant $H_r$ and $E_z$ are quasi-TM and are usually denoted as WGH or $E_z$ modes. Typically, modes are represented with two additional numbers $n$ and $p$, which describe the number of zero crossings (nodes) of the dominate field in the radial and axial directions (WG$_{m,n,p}$). Thus, the fundamental quasi-TE and quasi-TM mode families are written as WGE$_{m,0,0}$ and WGH$_{m,0,0}$ respectively as $n = p = 0$.

\subsection{Fields in the sapphire resonator}

\begin{figure}[htb]
\begin{center}
\includegraphics[width=6cm]{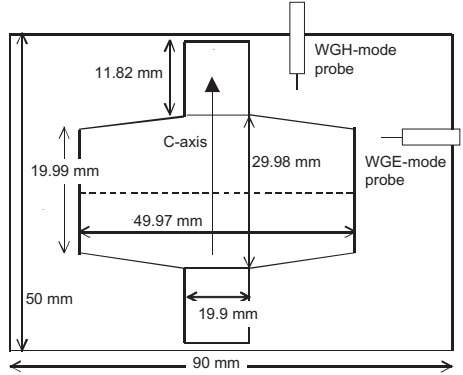}
\end{center}
\caption[]{Calculated dimensions of the sapphire cavity after contraction from room temperature to 4 K.}
\label{fig:Mike3}
\end{figure}

This subsection describes the techniques to calculate the fields in the uniaxial anisotropic sapphire resonator, with the crystal c-axis aligned with the z-axis \cite{TobMan}. A schematic of the sapphire resonator is shown in Fig. \ref{fig:Mike3}. Two techniques were used. The first was a Separation of Variables (SV) technique. The advantage was the ability to derive simple analytical expressions for all the field components, $E_z, H_z, E_\phi, H_\phi, E_r, H_r$. However, the technique is approximate, as it is impossible to calculate consistent analytical expressions for the six field components for an anisotropic right cylinder, with correct boundary conditions that satisfy Maxwell's equations. Furthermore, we have to make another approximation due to the sloped sides of our resonator (see Fig. \ref{fig:Mike3}). No doubt the most accurate technique is a numeric method, such as Finite Element (FE) analysis. However, for WG modes, field components calculated by FE and SV analysis are compared and shown to be close. Also, in Sect. 3.2, it is shown that calculations of the sensitivity to Lorentz violation in the Standard Model Extension, does not change significantly, whether we use the SV or FE results ($\approx 1\%$). Even if we use the pure WG mode approximation, sensitivity calculations are only over estimated by $\approx 5\%$.

\subsubsection{Separation of variables}

\begin{figure}[htb]
\begin{center}
\includegraphics[width=5cm]{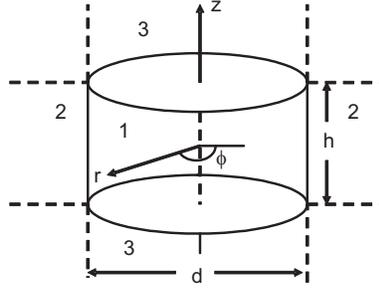}
\end{center}
\caption[]{Solution regions for separation of variables in cylindrical coordinates.}
\label{fig:Mike4}
\end{figure}

The Separation of Variables technique assumes the resonator is a perfect cylinder of uniaxial anisotropy with the crystal C-axis aligned along the cylinder axis and suspended in free space as shown in  Fig. \ref{fig:Mike4} \cite{TobMan}.

The $H_z$ magnetic field is symmetric along the z-axis (around z = 0). Given these conditions one can show that the solutions to Maxwell's equations yield:

\begin{eqnarray}
E_{z1} &=& i E_{z0} J_m(k_E r)\cos(m\phi)\sin(\beta z)\nonumber \\
E_{z2} &=& i E_{z02} K_m(k_{out} r)\cos(m\phi)\sin(\beta z)\nonumber \\
E_{z3} &=& i E_{z03} J_m(k_E r)\cos(m\phi)e^{-\alpha z} \label{zfields} \\ 
H_{z1} &=& H_{z0} J_m(k_H r)\sin(m\phi)\cos(\beta z)\nonumber \\
H_{z2} &=& H_{z02} K_m(k_{out} r)\sin(m\phi)\cos(\beta z)\nonumber \\
H_{z3} &=& H_{z03} J_m(k_H r)\sin(m\phi)e^{-\alpha z}\nonumber
\end{eqnarray}
where the constants and amplitudes are given in tables \ref{tab:Mike1} and \ref{tab:Mike2} below. One could also consider a region 4 in the corners where the field has $K_m(k_{out}r)$ and $e^{-\alpha z}$ dependence. However it is small, and in this work we ignore any field there. Also the azimuthal dependence is arbitrary, as long as we keep the $E$ and $H$ fields orthogonal either a $\cos(m\phi)$ or $\sin(m\phi)$ dependence could be assumed. In actual fact WG modes experimentally exhibit a doublet structure due to this degeneracy. In practice the degeneracy is split by a small fraction due to imperfections in cylindrical symmetry. Once the z components of the field are calculated, all the other components may be calculated from Maxwell's relationships in cylindrical co-ordinates (as shown in reference \cite{TobMan}). For example, the fields inside the sapphire (region 1 in Fig. \ref{fig:Mike4}) which concentrate over $98 \%$ of the total energy (c.f. Tab. \ref{tab:Mike3}) take the form

\begin{eqnarray}
H_z &=& H_{z0} J_m(k_H r)\sin(m\phi)\cos(\beta z)\nonumber \\
E_z &=& iE_{z0} J_m(k_E r)\cos(m\phi)\sin(\beta z) \label{Mollyfields} \\
H_r &=& h_{r0}\frac{1}{r}J_m(k_E r)\sin(m\phi)\sin(\beta z)+H_{r0}\left(J_{m-1}(k_H r)-J_{m+1}(k_H r)\right)\sin(m\phi)\sin(\beta z) \nonumber \\
E_r &=& ie_{r0}\frac{1}{r}J_m(k_H r)\cos(m\phi)\cos(\beta z)+iE_{r0}\left(J_{m-1}(k_E r)-J_{m+1}(k_E r)\right)\cos(m\phi)\cos(\beta z) \nonumber \\
H_\phi &=& h_{\phi 0}\frac{1}{r}J_m(k_H r)\cos(m\phi)\sin(\beta z)+H_{\phi 0}\left(J_{m-1}(k_E r)-J_{m+1}(k_E r)\right)\cos(m\phi)\sin(\beta z) \nonumber \\
E_\phi &=& ie_{\phi 0}\frac{1}{r}J_m(k_E r)\sin(m\phi)\cos(\beta z)+iE_{\phi 0}\left(J_{m-1}(k_H r)-J_{m+1}(k_H r)\right)\sin(m\phi)\cos(\beta z) \nonumber
\end{eqnarray}
where the amplitudes $(E_{z0}, h_{r0}, H_{r0}, e_{r0}, E_{r0}, h_{\phi 0}, H_{\phi 0}, e_{\phi 0}, E_{\phi 0})$ of the individual terms can be expressed as functions of the constants in Tab. \ref{tab:Mike1} and one of the amplitudes (e.g. $H_{z0}$). They are given in Tab. \ref{tab:Pete1} for our resonator.

An important point must be raised about matching boundary conditions to calculate the resonance frequency. The matching at the radial boundary generates hybrid solutions where both $H_z$ and $E_z$ field can co-exist. However, it is impossible to simultaneously match all field components on the axial boundary, so in some cases approximations must be made. However, this is a small effect as only some of the non-dominant components remain unmatched.

\subsubsection{Comparison of separation of variables with finite element analysis}

\begin{figure}[htb]
\begin{center}
\includegraphics[width=6cm]{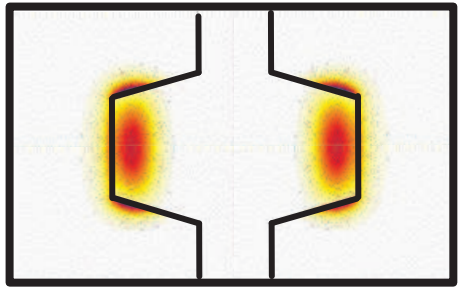}
\end{center}
\caption[]{Electric field density plot inside of the WGE$_{14,0,0}$ mode in the r-z plane.}
\label{fig:Mike5}
\end{figure}

Due to the sloped sides the SV analysis was checked using FE analysis \cite{Aubourg}. The calculated frequency was only out by 3 MHz compared to the measured frequency of 11.932 GHz. Typically up to 10 MHz discrepancy is normal for these types of calculations. The electric field density plot is shown in Fig. \ref{fig:Mike5}. It is apparent that we should be able to approximate the crystal as a right cylinder because the mode is mainly at the perimeter.

Given that the permittivity at 4 K is approximately $\epsilon_\| = 11.349$ and $\epsilon_\bot = 9.272$, to obtain a frequency close to the measured value and finite element calculation, an equivalent height of 21.45 mm is necessary, which gives a frequency of, $f = 11.931$ GHz close to the measured value of 11.932 GHz. The solution may be used to calculate the necessary parameters in (\ref{zfields}), which are summarized in Table \ref{tab:Mike1}.

\begin{table}[htb]
\caption{Calculated parameters (in $m^{-1}$) for the WGE$_{14,0,0}$ mode using the SV technique.}
\begin{center}
\begin{tabular}{cccccc}
\hline \hline
$m$ & $\alpha$ & $\beta$ & $k_H$ & $k_E$ & $k_{out}$\\
\hline
\ 14 \ & \ 707.420\ & \ 129.571\ & \ 750.314\ & \ 830.109\ & \ i213.868\ \\
\hline \hline
\end{tabular}
\end{center}
\label{tab:Mike1}
\end{table}

The amplitude of the fields may also be calculated. Because the WG mode is dominated by the $H_z$ component, we chose to relate the amplitude of all other components with respect to $H_{z0}$. Results are given in Table \ref{tab:Mike2}. To obtain $E_{z03}$ uniquely, the component of the tangential E field ($E_r$ and $E_\phi$ ) generated by $E_z$, must be assumed to be equal to zero to generate a consistent solution (i.e. only assume the $H_z$ component exists).

\begin{table}[htb]
\caption{Calculated amplitudes of the field components in (\ref{zfields}) with respect to $H_{z0}$.}
\begin{center}
\begin{tabular}{cccccc}
\hline \hline
$E_{z0}$ & $E_{z02}$ & $E_{z03}$ & $H_{z0}$ & $H_{z02}$ & $H_{z03}$\\
\hline
\ 19.4538$H_{z0}$ \ & \ 6.60905$\times 10^{-4}H_{z0}$\ & \ 428375$H_{z0}$\ & \ $H_{z0}$\ & \ -4.01697$\times 10^{-6}H_{z0}$\ & \ 355.373$H_{z0}$\ \\
\hline \hline
\end{tabular}
\end{center}
\label{tab:Mike2}
\end{table}

All other fields are then obtained from Mawxell relations. For example, inside the resonator (region 1 in Fig. \ref{fig:Mike4}) they are given by (\ref{Mollyfields}) with the amplitudes given in Tab. \ref{tab:Pete1}.

\begin{table}[htb]
\caption{Calculated amplitudes of the field components in (\ref{Mollyfields}) with respect to $H_{z0}$.}
\begin{center}
\begin{tabular}{ccccc}
\hline \hline
$E_{z0}$ & $h_{r0}$ & $H_{r0}$ & $e_{r0}$ & $E_{r0}$ \\
\hline
\ 19.4538\ $H_{z0}$ \ & \ 0.002978\ $H_{z0}$\ & \ -0.08634\ $H_{z0}$\ & \ -2.344\ $H_{z0}$\ & \ 1.858\ $H_{z0}$\\
\hline \hline
$h_{\phi 0}$ & $H_{\phi 0}$ & $e_{\phi 0}$ & $E_{\phi 0}$\\
\hline
-0.003222\ $H_{z0}$\ &\ 0.08828\ $H_{z0}$\ & \ -0.06268\ $H_{z0}$\ & \ 62.80\ $H_{z0}$\\
\hline\hline
\end{tabular}
\end{center}
\label{tab:Pete1}
\end{table}

Once the amplitude relationships are calculated the percentage of electric ($Pe_i$) and magnetic field ($Pm_i$) inside and outside the resonator may be calculated from:

\begin{equation}
Pe_i = \frac{\int_V \epsilon \vert E_i\vert^2 d^3x}{\int_V \epsilon ({\bf E}^* \cdot {\bf E}) d^3x} \hspace{1cm}
Pm_i = \frac{\int_V \mu \vert H_i\vert^2 d^3x}{\int_V \mu ({\bf H}^* \cdot {\bf H}) d^3x} 
\label{filfac}
\end{equation}
where the subscript $i$ refers to the component of the field ($r$, $\phi$ or $z$). The calculated filling factors using SV and FE are compared in Table \ref{tab:Mike3}.

\begin{table}[htb]
\caption{Calculated electric and magnetic filling factors in $r$, $\phi$, $z$ directions.}
\begin{center}
\begin{tabular}{ccccccc}
\hline \hline
Method & $Pe_z$ & $Pe_r$ & $Pe_\phi$ & $Pm_z$ & $Pm_r$ & $Pm_\phi$ \\
\hline
FE (in sapphire) \ & \ 0.007283\ & \ 0.8088\ & \ 0.1651\ & \ 0.9608\ & \ 0.005316\ & \ 0.02383\\
SV (in sapphire) \ & \ 0.006813\ & \ 0.8088\ & \ 0.1548\ & \ 0.9608\ & \ 0.005336\ & \ 0.02388\\
FE (in free space) \ & \ 0.0001642\ & \ 0.01061\ & \ 0.008057\ & \ 0.005594\ & \ 0.0007796\ & \ 0.003710\\
SV (in free space) \ & \ 0.01139\ & \ 0.0106\ & \ 0.007682\ & \ 0.00555\ & \ 0.0007885\ & \ 0.003616\\
\hline \hline
\end{tabular}
\end{center}
\label{tab:Mike3}
\end{table}

In general the mode is hybrid, and the sloped sides changes slightly the amplitude of the non dominant field components (all field components exist, although for this mode it is dominantly TE). This was verified by comparing the right cylinder to the one with sloped sides using FE analysis. To have an analytic solution that is consistent with the FE analysis, the amplitudes of some of the fields have been adjusted without changing the form of the solutions. This was achieved by leaving the amplitude of the dominant components ($H_z$ and $E_r$) the same, and changing all other components ($H_\phi$, $H_r$, $E_z$, $E_\phi$) by a unique amount to make the FE and SV solution of $Pe_r$ and $Pm_z$ the same within the sapphire. After this adjustment, it is clear that the solutions are similar. The normalizations factors used are shown in table \ref{tab:Mike4} and are close to unity.

\begin{table}[htb]
\caption{Amplitude normalization factors of the field components.}
\begin{center}
\begin{tabular}{ccccc}
\hline \hline
Field comp. & $H_\phi$ & $H_r$ & $E_z$ & $E_\phi$ \\
\hline
Norm. fact. \ & \ 0.82432\ & \ 0.73801\ & \ 0.569366\ & \ 1.00069\ \\
\hline \hline
\end{tabular}
\end{center}
\label{tab:Mike4}
\end{table}

The biggest discrepancy is with the $Pe_z$ component in free space. This is not surprising as it was the $E_{z03}$ amplitude in (\ref{zfields}), which was compromised with the SV technique. However, the technique allows us to obtain simple analytic solutions of the form (\ref{Mollyfields}) to the field for a WG mode resonator that can be useful when calculating the sensitivity to Lorentz violations within the framework of the Standard Model Extension \cite{KM} (see Sect. 3.2).

\section{Theoretical analysis}

Owing to their simplicity the kinematical frameworks of
\cite{Robertson,MaS} have been widely used to model and interpret
many previous experiments testing LLI
\cite{Wolf,Schiller,Brillet,Hils,Riis,WP} and we will follow that
route in Sect. 3.1. An analysis based on the more fundamental "Standard Model
Extension" (SME) \cite{Kosto1,KM} is then presented in Sect. 3.2.

By construction, kinematical frameworks do not allow for any
dynamical effects on the measurement apparatus. This implies that
in all inertial frames two clocks of different nature (e.g. based
on different atomic species) run at the same relative rate, and two
length standards made of different materials keep their relative
lengths. Coordinates are defined by the clocks and length
standards, and only the transformations between those coordinate
systems are modified. In general this leads to observable effects
on light propagation in moving frames but, by definition, to no
observable effects on clocks and length standards. In particular,
no attempt is made at explaining the underlying physics (e.g.
modified Maxwell and/or Dirac equations) that could lead to
Lorentz violating light propagation but leave e.g. atomic energy
levels unchanged. On the other hand dynamical frameworks (e.g. the
TH$\epsilon\mu$ formalism or the SME) in general use a modified
general Lagrangian that leads to modified Maxwell and Dirac
equations and hence to Lorentz violating light propagation and
atomic properties, which is why they are considered more
fundamental and more complete than the kinematical frameworks.
Furthermore, as shown in \cite{KM}, the SME is kept sufficiently
general to, in fact, encompass the kinematical frameworks and some
other dynamical frameworks (in particular the TH$\epsilon\mu$
formalism) as special cases, although there are no simple and
direct relationships between the respective parameters.

\subsection{The Robertson, Mansouri \& Sexl framework}

Kinematical frameworks for the description of Lorentz violation
have been pioneered by Robertson \cite{Robertson} and further
refined by Mansouri and Sexl \cite{MaS} and others. Fundamentally
the different versions of these frameworks are equivalent, and
relations between their parameters are readily obtained. As
mentioned above these frameworks postulate generalized
transformations between a preferred frame candidate $\Sigma(T,{\bf
X})$ and a moving frame ${\rm S}(t,{\bf x})$ where it is assumed
that in both frames coordinates are realized by identical
standards. We start from the transformations of \cite{MaS} (in
differential form) for the case where the velocity of $S$ as
measured in $\Sigma$ is along the positive X-axis, and assuming
Einstein synchronization in $S$ (we will be concerned with signal
travel times around closed loops so the choice of synchronization
convention can play no role):

\begin{equation}
dT = {1\over a}\left(dt+{vdx\over c^2}\right); dX = {dx\over
b}+{v\over a}\left(dt+{vdx\over c^2}\right); dY = {dy\over d}; dZ
= {dz\over d} \label{MStransf}
\end{equation}
with $c$ the velocity of light in vacuum in $\Sigma$. Using the
usual expansion of the three parameters $(a \approx
1+\alpha_{\mathrm{MS}}{v^2/c^2} + {\cal O}(4); b \approx
1+\beta_{\mathrm{MS}}{v^2/c^2} + {\cal O}(4); d \approx
1+\delta_{\mathrm{MS}}{v^2/c^2} + {\cal O}(4))$, setting
$c^2dT^2=dX^2+dY^2+dZ^2$ in $\Sigma$, and transforming according
to (\ref{MStransf}) we find the coordinate travel time of a light
signal in S:

\begin{equation}
dt={dl\over c}\left(1-\left(\beta_{\mathrm{MS}}
-\alpha_{\mathrm{MS}} -1 \right){v^2\over c^2} - \left({1\over
2}-\beta_{\mathrm{MS}} +\delta_{\mathrm{MS}} \right){\rm
sin}^2\theta{v^2\over c^2}\right)+{\cal O}(4) \label{MSc}
\end{equation}
where $dl = \sqrt{dx^2+dy^2+dz^2}$ and $\theta$ is the angle
between the direction of light propagation and the velocity {\bf
v} of S in $\Sigma$. In special relativity $\alpha_{\mathrm{MS}} =
-1/2; \beta_{\mathrm{MS}} = 1/2; \delta_{\mathrm{MS}} = 0$ and
(\ref{MStransf}) reduces to the usual Lorentz transformations.
Generally, the best candidate for $\Sigma$ is taken to be the
frame of the cosmic microwave background (CMB) \cite{Fixsen,Lubin}
with the velocity of the solar system in that frame taken as
$v_\odot \approx 377$ km/s, decl. $\approx -6.4 ^\circ $, $RA
\approx 11.2$h.

Michelson-Morley type experiments \cite{MM,Brillet} determine the
coefficient $P_{MM} = (1/2-\beta_{\mathrm{MS}}
+\delta_{\mathrm{MS}})$ of the direction dependent term. For many
years the most stringent limit on that parameter was $|P_{MM}|
\leq 5 \times 10^{-9}$ determined over 23 years ago in an
outstanding experiment \cite{Brillet}. Our experiment confirms
that result with roughly equivalent uncertainty $(2.2 \times
10^{-9})$. Recently an improvement to $|P_{MM}| \leq 1.5 \times
10^{-9}$ has been reported \cite{Muller}. Kennedy-Thorndike
experiments \cite{KT,Hils,Schiller} measure the coefficient
$P_{KT} = (\beta_{\mathrm{MS}} -\alpha_{\mathrm{MS}} -1)$ of the
velocity dependent term. The most stringent limit \cite{Schiller}
on $|P_{KT}|$ has been recently improved from \cite{Hils} by a
factor 3 to $|P_{KT}| \leq 2.1 \times 10^{-5}$. We improve this
result by a factor of 70 to $|P_{KT}| \leq 3.0 \times 10^{-7}$.
Finally clock comparison and Doppler experiments
\cite{Riis,WP,Grieser} measure $\alpha_{\mathrm{MS}}$. The most stringent result comes from the recent experiment of \cite{Saathoff}, limiting $|\alpha_{\mathrm{MS}} + 1/2|$ to $\leq 2.2 \times
10^{-7}$. The three types of experiments taken together then
completely characterize any deviation from Lorentz invariance in
this particular test theory, with present limits summarized in
Table \ref{MStab}.

\begin{table}
\caption{Present limits on Lorentz violating parameters in the
framework of \cite{MaS}}
\begin{center}
\renewcommand{\arraystretch}{1.4}
\setlength\tabcolsep{5pt}
\begin{tabular}{cccc}
\hline\hline\noalign{\smallskip}
Reference & $\alpha_{\mathrm{MS}} + 1/2$ & $1/2-\beta_{\mathrm{MS}} +\delta_{\mathrm{MS}}$ & $\beta_{\mathrm{MS}} -\alpha_{\mathrm{MS}} -1$ \\
\noalign{\smallskip} \hline \noalign{\smallskip}
\cite{Riis,Grieser,WP} & $\leq 8 \times 10^{-7}$ & - & -  \\
\cite{Saathoff} & $\leq 2.2 \times 10^{-7}$ & - & -  \\
\cite{Brillet} & - & $\leq 5 \times 10^{-9}$ & - \\
\cite{Muller} & - & $(2.2 \pm 1.5)\times 10^{-9}$ & - \\
\cite{Schiller} & - & - & $(1.9 \pm 2.1)\times 10^{-5}$ \\
our previous results \cite{Wolf} & - & $(1.5 \pm 4.2)\times 10^{-9}$ & $(-3.1 \pm 6.9)\times 10^{-7}$ \\
this work & - & $(1.2 \pm 2.2)\times 10^{-9}$ & $(1.6 \pm 3.0)\times 10^{-7}$ \\
\hline\hline
\end{tabular}
\end{center}
\label{MStab}
\end{table}

As described in Sect. 2 our cryogenic oscillator consists of a sapphire crystal of
cylindrical shape operating in a whispering gallery mode, and its coordinate frequency can be expressed
by equation (\ref{fWG}) where $t_c$ is the coordinate travel time of a
light signal around the circumference of the cylinder. Calculating $t_c$ from (\ref{MSc}) the relative
frequency difference between the sapphire oscillator and the
hydrogen maser (which, by definition, realizes coordinate time in
S \cite{masercom}) is

\begin{equation}
{\Delta\nu (t) \over \nu_0} = P_{KT}{v(t)^2\over c^2} +
P_{MM}{v(t)^2\over c^2}{1 \over 2\pi}\int_0^{2 \pi}{\rm
sin}^2\theta (t,\varphi ) d\varphi +{\cal O}(3) \label{MSdff}
\end{equation}
where $\nu_0$ is the unperturbed frequency, $v(t)$ is the (time dependent) speed
of the lab in $\Sigma$, and $\varphi$ is the azimuthal angle of
the light signal in the plane of the cylinder. The periodic time
dependence of $v$ and $\theta$ due to the rotation and orbital
motion of the Earth with respect to the CMB frame allow us to set
limits on the two parameters in (\ref{MSdff}) by adjusting the
periodic terms of appropriate frequency and phase (see \cite{Mike}
for calculations of similar effects for several types of
oscillator modes). Given the limited durations of our data sets
($\leq$ 15 days) the dominant periodic terms arise from the
Earth's rotation, so retaining only those we have ${\bf v}(t) =
{\bf u}+{\bf \omega} \times {\bf R}$ with ${\bf u}$ the velocity
of the solar system with respect to the CMB, ${\bf \omega}$ the
angular velocity of the Earth, and ${\bf R}$ the geocentric
position of the lab. We then find after some calculation.

\begin{equation}
\begin{array}{cl}
\Delta\nu / \nu_0 &= P_{KT}(H{\rm sin}\lambda )\\
\ & + P_{MM}(A{\rm cos}\lambda + B{\rm cos}(2\lambda)+C{\rm
sin}\lambda+D{\rm sin}\lambda{\rm cos}\lambda+E{\rm
sin}\lambda{\rm cos}(2\lambda))
\end{array}
\label{MSdff2}
\end{equation}
where $\lambda =\omega t + \phi$, and A-E and $\phi$ are constants
depending on the latitude and longitude of the lab $(\approx 48.7
^\circ$N and $2.33 ^\circ$E for Paris). Numerically $H \approx
-2.6 \times 10^{-9}$, $A \approx -8.8 \times 10^{-8}$, $B \approx
1.8 \times 10^{-7}$, C-E of order $10^{-9}$. We note that in
(\ref{MSdff2}) the dominant time variations of the two
combinations of parameters are in quadrature and at twice the
frequency which indicates that they should decorelate well in the
data analysis allowing a simultaneous determination of the two (as
confirmed by the correlation coefficients given in Sect. 4).
Adjusting this simplified model to our data we obtain results that
differ by less than 10\% from the results presented in Sect. 4
that were obtained using the complete model ((\ref{MSdff}) including the orbital motion of the Earth).

\subsection{The Standard Model Extension}

The fundamental theory of the Standard Model Extension (SME) as applied to electrodynamics is laid out in \cite{KM}. Here we summarise the main points relating to e-m cavity tests, and apply them to model our experiment.

The photon sector of the SME is described by a Lagrangian that takes the form

\begin{equation}
{\cal L} = -\frac{1}{4}F_{\mu \nu}F^{\mu \nu}+\frac{1}{2}(k_{AF})^\kappa\epsilon_{\kappa\lambda\mu\nu}A^\lambda F^{\mu\nu}
-\frac{1}{4}(k_F)_{\kappa\lambda\mu\nu}F^{\kappa \lambda}F^{\mu \nu}
\label{SMElag}
\end{equation}
where $F_{\mu\nu} \equiv \partial_\mu A_\nu - \partial_\nu A_\mu$. The first term is the usual Maxwell part while the second and third represent Lorentz violating contributions that depend on the parameters $k_{AF}$ and $k_F$. For most analysis the $k_{AF}$ parameter is set to 0 for theoretical reasons (c.f. \cite{KM}) but which is also well supported experimentally. The remaining dimensionless tensor $(k_F)_{\kappa\lambda\mu\nu}$ has a total of 19 independent components that need to be determined by experiment. Retaining only this term leads to Maxwell equations that take the familiar form but with $\bf D$ and $\bf H$ fields in vacuum defined by a six dimensional matrix equation

\begin{equation}
\left( \begin{array}{c}
{\bf D} \\
{\bf H} \end{array}\right)
 = \left( \begin{array}{c}
\epsilon_0(1+\kappa_{DE}) \\
\sqrt{\frac{\epsilon_0}{\mu_0}}\kappa_{HE} \end{array}
\begin{array}{c}
\sqrt{\frac{\epsilon_0}{\mu_0}}\kappa_{DB} \\
\mu_0^{-1}(1+\kappa_{HB}) \end{array} \right)
\left( \begin{array}{c}
{\bf E} \\
{\bf B} \end{array}\right)
\label{DHdef}
\end{equation}
where the $\kappa$ are $3\times 3$ matrices whose components are particular combinations of the $k_F$ tensor (c.f. equation (5) of \cite{KM}). Equation (\ref{DHdef}) indicates a useful analogy between the SME in vacuum and standard Maxwell equations in homogeneous anisotropic media. For the analysis of different experiments it turns out to be useful to introduce further combinations of the $\kappa$ matrices defined by:

\begin{eqnarray}
(\tilde{\kappa}_{e+})^{jk}&=&\frac{1}{2}(\kappa_{DE}+\kappa_{HB})^{jk}, \nonumber \\
(\tilde{\kappa}_{e-})^{jk}&=&\frac{1}{2}(\kappa_{DE}-\kappa_{HB})^{jk} - \frac{1}{3}\delta^{jk}(\kappa_{DE})^{ll}, \nonumber \\
(\tilde{\kappa}_{o+})^{jk}&=&\frac{1}{2}(\kappa_{DB}+\kappa_{HE})^{jk}, \nonumber \\
(\tilde{\kappa}_{o-})^{jk}&=&\frac{1}{2}(\kappa_{DB}-\kappa_{HE})^{jk}, \nonumber \\
(\tilde{\kappa}_{tr})^{jk}&=&\frac{1}{3}\delta^{jk}(\kappa_{DE})^{ll}.
\label{kappadef}
\end{eqnarray}

The first four of these equations define traceless $3 \times 3$ matrices, while the last defines a single coefficient. All $\tilde{\kappa}$ matrices are symmetric except $\tilde{\kappa}_{o+}$ which is antisymmetric. These characteristics leave a total of 19 independent coefficients of the $\tilde{\kappa}$. In general experimental results are quoted and compared using the $\tilde{\kappa}$ parameters rather than the original $k_F$ tensor components and this is the route we will follow in the present analysis.

The $k_F$ tensor in (\ref{SMElag}), and consequently the $\kappa$ tensors in (\ref{DHdef}) and (\ref{kappadef}), are frame dependent and consequently vary as a function of the coordinate system chosen to analyse a given experiment. In principle they may be constant and non-zero in any frame (e.g. the cavity frame or the lab frame). However, any non-zero values are expected to arise from Planck-scale effects in the early Universe. Therefore the components of $k_F$ should be constant in a cosmological frame (e.g. the one defined by the CMB radiation) or any frame that moves with a constant velocity and shows no rotation with respect to the cosmological one. Consequently the conventionally chosen frame to analyse and compare experiments in the SME is a sun-centred, non-rotating frame as defined in \cite{KM}. The general proceedure is to calculate the perturbation of the resonator frequency as a function of the unperturbed $E_0$ and $B_0$ fields and $\kappa$ tensors in the lab frame and then to transform the $\kappa$ tensors to the conventional sun-centred frame. This transformation will introduce a time variation of the frequency related to the movement of the lab with respect to the sun-centred frame (typically introducing time variations of sidereal and semi-sidereal periods for an Earth fixed experiment).

In \cite{KM} the authors derive an expression for the perturbed frequency of a resonator of the form 

\begin{eqnarray}
\frac{\Delta\nu}{\nu_0} = &-&\frac{1}{\langle U \rangle} \int_V d^3x \left(\epsilon_0 {\bf E_0}^*\cdot \kappa_{DE}\cdot {\bf E_0} - \mu_0^{-1} {\bf B_0}^*\cdot \kappa_{HB}\cdot {\bf B_0} \right. \\
&+& \left. 2{\rm Re}(\sqrt{\frac{\epsilon_0}{\mu_0}}{\bf E_0}^*\cdot \kappa_{DB}\cdot {\bf B_0}) \right) \nonumber
\label{KM34}
\end{eqnarray}
where ${\bf B_0}, {\bf H_0}, {\bf E_0}, {\bf D_0}$ are the unperturbed (standard Maxwell) fields and $\langle U \rangle = \int_V d^3x ({\bf E_0}\cdot {\bf D_0}^*+{\bf B_0}\cdot {\bf H_0}^*)$. This expression can be applied directly to our resonator using the fields calculated in Sect. 2.2.

The resonator is placed in the lab with its symmetry axis along the vertical. So applying (\ref{KM34}) in the lab frame (z-axis vertical upwards, x-axis pointing south), with the fields calculated as described in Sect. 2.2, we obtain an expression for the frequency variation of the resonator

\begin{eqnarray}
\frac{\Delta\nu}{\nu_0}&=& ({\cal M}_{DE})_{lab}^{xx}\left((\kappa_{DE})_{lab}^{xx}+(\kappa_{DE})_{lab}^{yy}\right)+({\cal M}_{DE})_{lab}^{zz}(\kappa_{DE})_{lab}^{zz} \nonumber \\
&+& ({\cal M}_{HB})_{lab}^{xx}\left((\kappa_{HB})_{lab}^{xx}+(\kappa_{HB})_{lab}^{yy}\right)+({\cal M}_{HB})_{lab}^{zz}(\kappa_{HB})_{lab}^{zz}
\label{dffM}
\end{eqnarray}
with the ${\cal M}_{lab}$ components given in Tab. \ref{tab:Mike5}. To obtain the values in Tab. \ref{tab:Mike5} we take into account the fields inside the resonator (e.g. (\ref{Mollyfields})) and outside ($\leq 2\%$ of the energy). We note that when using coordinates in which the $\kappa_{lab}^{ij}$ components are constants over the volume of the cylinder (e.g. cartesian coordinates), the integrals in (\ref{KM34}) are equivalent to filling factor integrals (c.f. (\ref{filfac})) for the diagonal terms. This simplifies the calculation of the ${\cal M}_{lab}^{ll}$ components when using numerical techniques, as simple algorithims exist to calculate filling factors without explicitly calculating the field \cite{Krupka2}.

\begin{table}[htb]
\caption{${\cal M}_{lab}$ components calculated using the different techniques for the determination of the fields inside the resonator described in Sect. 2.2.}
\begin{center}
\begin{tabular}{cccc}
\hline \hline
${\cal M}_{lab}^{ij}$ & FE & SV & Pure WG \\
\hline
$({\cal M}_{DE})_{lab}^{xx}$ \ & \ -0.03093\ & \ -0.0355\ & \ -0.02696\ \\
$({\cal M}_{DE})_{lab}^{zz}$ \ & \ -0.0004030\ & \ -0.005996\ & \ 0.00\ \\
$({\cal M}_{HB})_{lab}^{xx}$ \ & \ 0.008408\ & \ 0.008405\ & \ 0.00\ \\
$({\cal M}_{HB})_{lab}^{zz}$ \ & \ 0.4832\ & \ 0.4832\ & \ 0.50\ \\
\hline \hline
\end{tabular}
\end{center}
\label{tab:Mike5}
\end{table}

The last step is to transform the $\kappa$ tensors in (\ref{dffM}) to the conventional sun-centred frame using the explicit transformations provided in \cite{KM}, and to express the result in terms of the $\tilde{\kappa}$ tensors of (\ref{kappadef}). We obtain

\begin{eqnarray}
\frac{\Delta\nu}{\nu} &=& K + C_A\cos(\Omega T) + S_A\sin(\Omega T) \nonumber \\
&+& C_\omega\cos(\omega T_\oplus) + S_\omega\sin(\omega T_\oplus) \nonumber \\
&+& C_{2\omega}\cos(2\omega T_\oplus) + S_{2\omega}\sin(2\omega T_\oplus) \label{dffSME}\\
&+& C_{\omega +\Omega}\cos((\omega+\Omega)T_\oplus+\varphi) + S_{\omega +\Omega}\sin((\omega+\Omega)T_\oplus+\varphi) \nonumber \\
&+& C_{\omega -\Omega}\cos((\omega-\Omega)T_\oplus-\varphi) + S_{\omega -\Omega}\sin(((\omega-\Omega)T_\oplus-\varphi) \nonumber \\
&+& C_{2\omega +\Omega}\cos((2\omega+\Omega)T_\oplus+\varphi) + S_{2\omega +\Omega}\sin((2\omega+\Omega)T_\oplus+\varphi) \nonumber \\
&+& C_{2\omega -\Omega}\cos((2\omega-\Omega)T_\oplus-\varphi) + S_{2\omega -\Omega}\sin((2\omega-\Omega)T_\oplus-\varphi) \nonumber 
\end{eqnarray}
where $\omega$ and $\Omega$ are the sidereal and anual angular frequencies ($\omega = 7.292115 \times 10^{-5}$ rad/s, $\Omega = 1.991 \times 10^{-7}$ rad/s), $T_\oplus$ is the time since a coincidence of the lab y axis with the sun-centred Y axis (as defined in \cite{KM}), $T$ is the time since a spring equinox and $\varphi$ is a constant given by $\varphi = \Omega(T-T_\oplus)$. 

The $\tilde{\kappa}_{e+}$ and $\tilde{\kappa}_{o-}$ tensors have been determined to $\leq 2 \times 10^{-32}$ by astrophysical tests \cite{KM}, orders of magnitude below what we can hope to achieve in resonator experiments. Consequently we set those parameters to zero, obtaining the $C$ and $S$ coefficients of (\ref{dffSME}) given in Tab. \ref{tab:CwSw}. They involve 4 combinations of the 5 independent components of $\tilde{\kappa}_{e-}$ and all three independent components of $\tilde{\kappa}_{o+}$, but do not involve $\tilde{\kappa}_{tr}$. So our experiment is sensitive to 7 of the 9 parameters not determined by astrophysical tests. The numerical values in Tab. \ref{tab:CwSw} were obtained using the FE results in Tab. \ref{tab:Mike5}. Similar values are obtained when using the SV or 'pure WG' results, with the differences not exceeding 1\% and 5\% respectively. This small discrepancy is due to the fact that the dominant term ($({\cal M}_{HB})_{lab}^{zz}$) in Tab. \ref{tab:Mike5} is almost unaffected by the technique used. However, after transformation to the sun centred frame and to the $\tilde{\kappa}$ tensors it is that term which dominates the sensitivity to the $\tilde{\kappa}$ components.

Present limits on the 7 parameters are summarised in Tab. \ref{tab:SMElimits}. At least one year of regular data is required to be able to decorrelate the $\omega$ and $2\omega$ frequencies from the $\omega \pm \Omega$ and $2\omega \pm \Omega$ ones. The data of our experiment that we present here (see next section) are not yet sufficient for that purpose (correlation coefficients are still $\approx 0.4$), so the next section only presents the results in the kinematical framework of Sect. 3.1. However, the experiment is still running, and we expect to publish our results on the SME test by early 2004.

\begin{table}[htb]
\caption{Coefficients in (\ref{dffSME}). The constant $K$ and the annual coefficients ($C_A$ and $S_A$) are not given as our experiment is insensitive to those terms.}
\begin{center}
\begin{tabular}{ccc}
\hline \hline
frequency & $C$ & $S$ \\
\hline
$\omega$ & $-0.4402 \ \tilde{\kappa}_{e-}^{13}+(1.066\times 10^{-6}) \ \tilde{\kappa}_{o+}^{13}$\hspace{3mm}  &\hspace{3mm}  $-0.4402 \ \tilde{\kappa}_{e-}^{23}+(1.066\times 10^{-6}) \ \tilde{\kappa}_{o+}^{23}$\\
$2\omega$ & $-0.09625 \ (\tilde{\kappa}_{e-}^{11}-\tilde{\kappa}_{e-}^{22})$\hspace{3mm}  &\hspace{3mm} $-0.1925 \ \tilde{\kappa}_{e-}^{12}$ \\
$\omega+\Omega$ & $-(8.6405\times 10^{-6})\ \tilde{\kappa}_{o+}^{23}$\hspace{3mm}  &\hspace{3mm} $(8.6405\times 10^{-6})\ \tilde{\kappa}_{o+}^{13}+(1.779\times 10^{-6})\ \tilde{\kappa}_{o+}^{12}$\\
$\omega-\Omega$ & $-(8.6405\times 10^{-6})\ \tilde{\kappa}_{o+}^{23}$\hspace{3mm}  &\hspace{3mm} $(8.6405\times 10^{-6})\ \tilde{\kappa}_{o+}^{13}-(4.196\times 10^{-5})\ \tilde{\kappa}_{o+}^{12}$\\
$2\omega+\Omega$ & $(7.780\times 10^{-7})\ \tilde{\kappa}_{o+}^{13}$\hspace{3mm}  &\hspace{3mm} $(7.780\times 10^{-7})\ \tilde{\kappa}_{o+}^{23}$\\
$2\omega-\Omega$ & $-(1.834\times 10^{-5})\ \tilde{\kappa}_{o+}^{13}$\hspace{3mm}  &\hspace{3mm} $-(1.834\times 10^{-5})\ \tilde{\kappa}_{o+}^{23}$\\ 
\hline \hline
\end{tabular}
\end{center}
\label{tab:CwSw}
\end{table}

\begin{table}[htb]
\caption{Present limits (1$\sigma$ uncertainties) on the 7 SME parameters our experiment is sensitive to, as determined in \cite{Muller}. Values are given in $10^{-15}$ for $\tilde{\kappa}_{e-}$ and $10^{-11}$ for $\tilde{\kappa}_{o+}$.}
\begin{center}
\begin{tabular}{ccccccc}
\hline \hline
$\tilde{\kappa}_{e-}^{12}$ &\ $\tilde{\kappa}_{e-}^{13}$ & $\tilde{\kappa}_{e-}^{23}$ &\ $\tilde{\kappa}_{e-}^{11}-\tilde{\kappa}_{e-}^{22}$\hspace{5mm} &\ \ $\tilde{\kappa}_{o+}^{12}$ & $\tilde{\kappa}_{o+}^{13}$ & $\tilde{\kappa}_{o+}^{23}$\\
\hline
$1.7 \pm 2.6$\ &\ $-6.3 \pm 12.4$\ &\ $3.6 \pm 9.0$\ &\ $8.9 \pm 4.9$\hspace{5mm} &\ $-14 \pm 14$\ &\ $1.2 \pm 2.6$\ &\ $-0.1 \pm 2.7$\\ 
\hline \hline
\end{tabular}
\end{center}
\label{tab:SMElimits}
\end{table}

We conclude this section by a short discussion of the effects of non-zero values of the SME parameters on our frequency reference (the hydrogen maser) and the crystal structure of the sapphire (and hence the resonator shape). Such effects would add to the direct effect on the electromagnetic fields and may therefore lead to overall cancellation or enhancement of the sensitivities calculated above. However, it was shown in \cite{Muller2} that the effect on the sapphire crystal amounts to only a few percent of the direct effect on the fields, and \cite{KL} show that the hydrogen $m_F=0 \to m'_F=0$ clock transition is not affected to first order. Hence the total effect is dominated by the Lorentz violating properties of the electromagnetic fields inside the resonator and well described (to a few percent) by the model derived above. 

\section{Experimental results}
As mentioned in the previous section, the experimental data presented here does not yet allow a complete decorrelation of the different parameters in the SME model, so we concentrate in this section on the experimental results in the kinematic (RMS) framework presented in section 3.1.

The cryogenic Sapphire Oscillator (CSO) is an active system oscillating at the resonant frequency (i.e. a classical loop oscillator which amplifies and re-injects the "natural" resonator signal). Additionally the signal is locked to the resonance using the Pound-Drever technique (modulation at $\approx$ 80 kHz). The incident power is stabilised in the cryogenic environment and the spurious AM modulation is minimised using a servo loop. To minimise temperature sensitivity the resonator is heated (inside the 4 K environment) and stabilised to the temperature turning point ($\approx$ 6 K) of the resonator frequency which arises due to paramagnetic impurities in the sapphire. Under these conditions the loaded quality factor of the resonator is slightly below $10^9$. The resonator is kept permanently at cryogenic temperatures, with helium refills taking place about every 20 - 25 days.

The CSO is compared to a commercial (Datum Inc.) active hydrogen maser whose frequency is also regularly compared to caesium and rubidium atomic fountain clocks in the laboratory \cite{Bize}. The CSO resonant frequency at 11.932 GHz is compared to the 100 MHz output of the hydrogen maser. The maser signal is multiplied up to 12 GHz of which the CSO signal is subtracted. The  remaining $\approx$ 67 MHz signal is mixed to a synthesizer signal at the same frequency and the low frequency beat at $\approx$ 64 Hz is counted, giving access to the frequency difference between the maser and the CSO. The instability of the comparison chain has been measured at $\leq 2 \times 10^{-14}\tau^{-1}$, with long term instabilities dominated by temperature variations, but not exceeding $10^{-16}$.

Since September 2002 we are taking continuous temperature measurements on top of the CSO dewar and behind the electronics rack. Starting January 2003 we have implemented an active temperature control of the CSO room and changed some of the electronics. As a result the diurnal and semi-diurnal temperature variations during measurement runs ($\approx$ 2 weeks) were greatly reduced to less than $0.025^\circ$ C in amplitude (best case), and longer and more reliable data sets became available.

Our previously published results \cite{Wolf} are based on data sets taken between Nov. 2001 and Sep. 2002. Most of that data (except the last data set) was taken before implementation of permanent temperature control. As a result the uncertainties in \cite{Wolf} were dominated by the systematic effects from temperature variations. Here we use only data that was permanently temperature controlled, 13 data sets in total spanning Sept. 2002 to Aug. 2003, of differing lengths (5 to 16 days, 140 days in total), thereby significantly reducing the uncertainties from systematic effects. The sampling time for all data sets was $100$ s except two data sets with $\tau_0 = 12$ s. To make the data more manageable we first average all points to $\tau_0 = 2500$ s. For the data analysis we simultaneously adjust an offset and a rate (natural frequency drift, typically $\approx 1.7 \times 10^{-18}$ s$^{-1}$) per data set and the two parameters of the model (\ref{MSdff}). In the model (\ref{MSdff}) we take into account the rotation of the Earth and the Earth's orbital motion, the latter contributing little as any constant or linear terms over the durations of the individual data sets are absorbed by the adjusted offsets and rates.

When carrying out an ordinary least squares (OLS) adjustment we note that the residuals have a significantly non-white behavior. The power spectral density (PSD) of the residuals when fitted with a power law model of the form $S_y(f)=kf^\mu$ yields typically $\mu \approx -1.5$ to $-2$. In the presence of non-white noise OLS is not the optimal regression method \cite{lss,Draper} as it can lead to significant underestimation of the parameter uncertainties \cite{lss}.

\begin{figure}[b]
\begin{center}
\includegraphics[width=9cm]{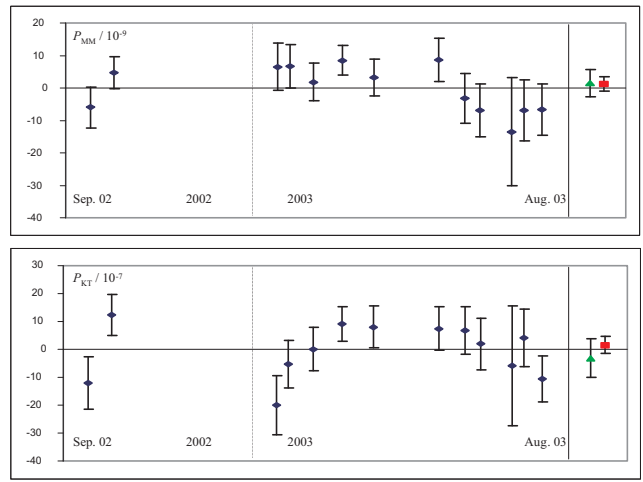}
\end{center}
\caption[]{Values of the two parameters ($P_{KT}$ and $P_{MM}$) from a fit to each individual data set (blue diamonds) and a global fit to all the data (red squares). For comparison our previously published results \cite{Wolf} are also shown (green triangles). The error bars indicate the combined uncertainties from statistics and systematic effects.}
\label{MMKTfig2}
\end{figure}

An alternative method is weighted least squares (WLS) \cite{Draper} which allows one to account for non-random noise processes in the original data by pre-multiplying both sides of the design equation (our equation (\ref{MSdff}) plus the offsets and rates) by a weighting matrix containing off diagonal elements. To determine these off diagonal terms we first carry out OLS and adjust the $S_y(f)=kf^\mu$ power law model to the PSD of the post-fit residuals determining a value of $\mu$ for each data set. We then use these $\mu$ values to construct a weighting matrix following the method of fractional differencing described, for example, in \cite{lss}. Figure \ref{MMKTfig2} shows the resulting values of the two parameters ($P_{KT}$ and $P_{MM}$) for each individual data set. A global WLS fit of the two parameters and the 13 offsets and drifts yields $P_{MM} = (1.2\pm 1.9) \times 10^{-9}$ and $P_{KT} = (1.6\pm 2.3) \times 10^{-7}$ ($1\sigma$ uncertainties), with the correlation coefficient between the two parameters less than 0.01 and all other correlation coefficients $< 0.06$. The distribution of the 13 individual values around the ones obtained from the global fit is well compatible with a normal distribution ($\chi^2$ = 10.7 and $\chi^2$ = 14.6 for $P_{MM}$ and $P_{KT}$ respectively).

Systematic effects at diurnal or semi-diurnal frequencies with the appropriate phase could mask a putative sidereal signal. The statistical uncertainties of $P_{MM}$ and $P_{KT}$ obtained from the WLS fit above correspond to sidereal and semi-sidereal terms (from (\ref{MSdff2})) of $\approx 7 \times 10^{-16}$ and $\approx 4 \times 10^{-16}$ respectively so any systematic effects exceeding these limits need to be taken into account in the final uncertainty. We expect the main contributions to such effects to arise from temperature, pressure and magnetic field variations that would affect the hydrogen maser, the CSO and the associated electronics, and from tilt variations of the CSO which are known to affect its frequency.

To estimate the tilt sensitivity we have intentionally tilted the oscillator by $\approx$ 5 mrad off its average position which led to relative frequency variations of $\approx 3 \times 10^{-13}$ from which we deduce a tilt sensitivity of $\approx 6 \times 10^{-17} \mu$rad$^{-1}$. Measured tilt variations in the lab at diurnal and semi-diurnal periods were measured at 4.6 $\mu$rad and 1.6 $\mu$rad respectively which leads to frequency variations that do not exceed $3 \times 10^{-16}$ and $1 \times 10^{-16}$ respectively and are therefore negligible with respect to the statistical uncertainties.

In December 2002 we implemented an active temperature stabilization inside an isolated volume ($\approx 15 {\rm m}^3$) that included the CSO and all the associated electronics. The temperature was measured continously in two fixed locations (behind the electronics rack and on top of the dewar). For the best data sets the measured temperature variations did not exceed 0.02/0.01 $^\circ$C in amplitude for the diurnal and semi-diurnal components. In the worst cases (the two 2002 data sets and some data sets taken during a partial air conditioning failure) the measured temperature variations could reach 0.26/0.08 $^\circ$C. When intentionally heating and cooling the CSO lab by $\approx 3^\circ$C we see frequency variations of $\approx 5 \times 10^{-15}$ per $^\circ$C. This is also confirmed when we induce a large sinusoidal temperature variation ($\approx 1.5 ^\circ$C amplitude). Using this we can calculate a value for temperature induced frequency variations at diurnal and semi-diurnal frequencies for each data set, obtaining values that range from $\approx 5 \times 10^{-17}$ to $\approx 1.3 \times 10^{-15}$.

The hydrogen maser is kept in a dedicated, environmentally controlled clock room. Measurements of magnetic field, temperature and atmospheric pressure in that room and the maser sensitivities as specified by the manufacturer allow us to exclude any systematic effects on the maser frequency that would exceed the statistical uncertainties above and the systematic uncertainties from temperature variations in the CSO lab.

Our final uncertainties (the error bars in Fig. \ref{MMKTfig2}) are the quadratic sums of the statistical uncertainties from the WLS adjustment for each data set and the systematic uncertainties calculated for each data set from (\ref{MSdff2}) and the measured temperature variations. For the global adjustment we average the systematic uncertainties from the individual data sets obtaining $\pm 1.2 \times 10^{-9}$ on $P_{MM}$ and $\pm 1.9 \times 10^{-7}$ on $P_{KT}$. Adding these quadratically to the WLS statistical uncertainties of the global adjustment we obtain as our final result $P_{MM} = (1.2\pm 2.2) \times 10^{-9}$ and $P_{KT} = (1.6\pm 3.0) \times 10^{-7}$ ($1\sigma$ uncertainties).

\section{Conclusion}

We have presented a detailed description of whispering gallery modes in cylindrical resonators. This includes analytical and numerical calculations of electro-magnetic fields and energy filling factors for the sapphire resonator used in our experiment. It was shown that the different calculations gave similar results (differences at the \% level). We then applied those results to model Lorentz violating frequency shifts of the resonator in two different theoretical frameworks: the kinematical framework of Robertson, Mansouri \& Sexl (RMS), and the standard model extension (SME) developed by Kostelecky and co-workers. In both cases we obtain explicit expressions ((\ref{MSdff}) and (\ref{dffSME})) for the fractional frequency variation of our resonator as a function of the Earth's rotational and orbital angular velocities and of the parameters describing Lorentz violation in RMS or the SME. We show that the experimental sensitivity to the SME parameters is not affected by more than 1 \% when using the analytical or numerical method to calculate the resonator fields.

Experimental results were given for the RMS test, showing that our experiment simultaneously constrains two combinations of the three parameters of the Mansouri and Sexl test theory (previously measured individually by Michelson-Morley and Kennedy-Thorndike experiments). We obtain $\delta_{\mathrm{MS}} - \beta_{\mathrm{MS}} + 1/2 = 1.2(1.9)(1.2) \times 10^{-9}$ which is of the same order as the best previous results \cite{Muller,Brillet}, and $\beta_{\mathrm{MS}} - \alpha_{\mathrm{MS}} - 1 = 1.6(2.3)(1.9)\times 10^{-7}$ which improves the best previous limit \cite{Schiller} by a factor of 70 (the first bracket indicates the $1\sigma$ uncertainty from statistics the second from systematic effects). We improve our own previous results \cite{Wolf} by about a factor 2 due to more and longer data sets and to improved temperature control of the experiment (see Tab. \ref{MStab} for a summary of present limits). We note that our value on $\delta_{\mathrm{MS}} - \beta_{\mathrm{MS}} + 1/2$ is compatible with the slightly significant recent result of \cite{Muller} who obtained $\delta_{\mathrm{MS}} - \beta_{\mathrm{MS}} + 1/2 = (2.2 \pm 1.5)\times 10^{-9}$.

As a result of our experiment the Lorentz transformations are confirmed in the RMS framework with an overall uncertainty of $\leq 3 \times 10^{-7}$ limited by our determination of $\beta_{\mathrm{MS}} - \alpha_{\mathrm{MS}} - 1$ and the recent limit \cite{Saathoff} of $2.2 \times 10^{-7}$ on the determination of $\alpha_{\mathrm{MS}}$. The latter is likely to improve in the coming years by experiments such as ACES (Atomic
Clock Ensemble in Space \cite{ACES}) that will compare ground
clocks to clocks on the international space station aiming at a $10^{-8}$ measurement of $\alpha_{\mathrm{MS}}$.

Concerning the SME we do not present experimental results as our data is not yet sufficient to decorrelate all SME parameters. However, the experiment is still running and we expect to have enough data by early 2004 to obtain unambiguous valus of all SME parameters. Nonetheless, we can already note that our experiment is complementary with respect to the best previous determination of those parameters \cite{Muller} by looking at tables \ref{tab:CwSw} and \ref{tab:SMElimits}. From Tab. \ref{tab:SMElimits} we see that the most accurate values are those for $\tilde{\kappa}_{e-}^{12}$ and $\tilde{\kappa}_{e-}^{11} - \tilde{\kappa}_{e-}^{22}$, the two parameters our experiment is less sensitive to (c.f. second line of Tab. \ref{tab:CwSw}) whereas our best sensitivity is to $\tilde{\kappa}_{e-}^{13}$ and $\tilde{\kappa}_{e-}^{23}$ (first line of Tab. \ref{tab:CwSw}) which are less well determined in \cite{Muller}. Given our uncertainties at sidereal and semi-sidereal periods of $\approx 1 \times 10^{-15}$ and $\approx 5 \times 10^{-16}$ respectively (c.f. Sect. 4) we expect to obtain uncertainties of about $2 \times 10^{-15}$ on $\tilde{\kappa}_{e-}^{13}$, $\tilde{\kappa}_{e-}^{23}$ and about $3 - 5 \times 10^{-15}$ on $\tilde{\kappa}_{e-}^{12}$, $\tilde{\kappa}_{e-}^{11} - \tilde{\kappa}_{e-}^{22}$. Those results, if achieved, will compare favourably to the present limits of Tab. \ref{tab:SMElimits}.

For the future we do not expect significant improvements using our present experimental setup, due to the already relatively long total data span and to expected systematic limits from both, the hydrogen maser and the sapphire resonator in the low $10^{-16}$ region. Significant improvements in the near future are more likely to come from new proposals, for example, using two orthogonal resonators or two orthogonal modes in the same sapphire resonator placed on a rotating platform \cite{Mike}. Such a set-up is likely to improve the tests of LLI by several orders of magnitude as the relevant time variations will now be at the rotation frequency ($\approx 0.01 - 0.1$ Hz) which is the range in which such resonators are the most stable ($\approx$ 100 fold better stability). Additionally many systematic effects are likely to cancel between the two orthogonal oscillators or modes and the remaining ones are likely to be less coupled to the rotation frequency than to the sidereal frequencies used in our experiment. Ultimately, it has been proposed \cite{Lammerzahl2001} to conduct these tests on board an Earth orbiting satellite, again with a potential gain of several orders of magnitudes over current limits.

\vspace{1cm}
{\bf \large Acknowledgments}
Helpful discussions (on Sect. 3.2 in particular) with Alan Kostelecky are gratefully acknowledged as well as financial support by the Australian Research Council and CNES. We also thank Pierre Guillon and Dominique Cros for access to Finite Element software provided by IRCOM at the University of Limoges. P.W. was supported by CNES research grant 793/02/CNES/4800000078.


\begin{thebibliography}{99}
\bibitem{Will} Will C.M., {\it Theory and Experiment in Gravitational Physics, revised edition}, Cambridge U. Press, (1993). 
\bibitem{KostoSam} Kostelecky V.A. and Samuel S., Phys. Rev. {\bf D39}, 683, (1989).
\bibitem{Damour1} Damour T., gr-qc/9711060 (1997).
\bibitem{Damour2} Damour T. and Polyakov A.M., Nucl.Phys. {\bf B423}, 532, (1994).
\bibitem{Kosto1} Colladay D. and Kostelecky V.A., Phys. Rev. {\bf D55}, 6760, (1997), Colladay D. and Kostelecky V.A., Phys. Rev. {\bf D58}, 116002 (1998).
\bibitem{Kosto2} Bluhm R. et al., Phys. Rev. Lett. {\bf 88}, 9, 090801, (2002).
\bibitem{Robertson} Robertson H.P., Rev. Mod. Phys. {\bf 21}, 378 (1949).
\bibitem{MaS} Mansouri R. and Sexl R.U., Gen. Rel. Grav. {\bf 8}, 497, 515, 809, (1977).
\bibitem{LightLee} Lightman A.P. and Lee D.L., Phys. Rev. {\bf D8}, 2, 364, (1973).
\bibitem{Blanchet} Blanchet L., Phys. Rev. Lett. {\bf 69}, 4, 559, (1992).
\bibitem{Ni} Ni W.-T., Phys. Rev. Lett. {\bf 38}, 301, (1977).
\bibitem{KM} Kostelecky A.V. and Mewes M., Phys. Rev. {\bf D66}, 056005, (2002).
\bibitem{Wolf} Wolf P. et al., Phys. Rev. Lett. {\bf 90}, 6, 060402, (2003).
\bibitem{Schiller} Braxmaier C. et al., Phys. Rev. Lett. {\bf 88}, 1, 010401, (2002).
\bibitem{Muller} M\"uller H. et al., Phys. Rev. Lett. {\bf 91}, 2, 020401, (2003).
\bibitem{Brillet} Brillet A. and Hall J.L., Phys. Rev. Lett. {\bf 42}, 9, 549, (1979).
\bibitem{KT} Kennedy R.J. and Thorndike E.M., Phys. Rev. {\bf B42}, 400, (1932).
\bibitem{Hils} Hils D. and Hall J.L., Phys. Rev. Lett., {\bf 64}, 15, 1697, (1990).
\bibitem{TobMan} M. E. Tobar and A. G. Mann, IEEE Transactions on Microwave Theory and Techniques, {\bf 39}, 2077, (1991).
\bibitem{Cros} D. Cros and P. Guillon, IEEE Transactions on Microwave Theory and Techniques, {\bf 38}, 1667, (1990).
\bibitem{Guillon}	P. Guillon and X. Jiao, Proc. IEE, Part H, {\bf 134}, (1987).
\bibitem{Krupka} J. Krupka, D. Cros, M. Aubourg, and P. Guillon, IEEE Transactions on Microwave Theory and Techniques, {\bf 41},56, (1994).
\bibitem{Beyer} S. Schiller and R. L. Beyer, Optics Lett., {\bf 16}, 1138, (1991).
\bibitem{TobAns} M.E. Tobar, J.D. Anstie, J.G. Hartnett, IEEE Trans. UFFC, {\bf 50}, 11, 1407, (2003).
\bibitem{Aubourg} M. Aubourg and P. Guillon, Journal of Electromagnetic Waves and Application, {\bf 45}, 71, (1991). 
\bibitem{Riis} Riis E. et al., Phys. Rev. Lett. {\bf 60}, 81, (1988).
\bibitem{WP} Wolf P. and Petit G., Phys. Rev. {\bf A56}, 6, 4405, (1997).
\bibitem{Fixsen} Fixsen D.J. et al., Phys. Rev. Lett. {\bf 50}, 620, (1983).
\bibitem{Lubin} Lubin et al., Phys. Rev. Lett. {\bf 50}, 616, (1983).
\bibitem{MM} Michelson A.A. and Morley E.W., Am. J. Sci., {\bf 34}, 333, (1887).
\bibitem{Grieser} Grieser R. et al., Appl. Phys. {\bf B59}, 127, (1994).
\bibitem{Saathoff} Saathoff G., et al., Phys. Rev. Lett. {\bf 91}, 19, 190403, (2003).
\bibitem{masercom} We assume here that local position invariance is sufficiently verified so that the variation of the maser frequency due to the diurnal variation of the local gravitational potential is negligible. Indeed the results of \cite{Bauch} imply that such variations should not exceed 2 parts in $10^{-17}$ which is significantly below our noise level.
\bibitem{Bauch} Bauch A., Weyers S., Phys. Rev. {\bf D65}, 081101, (2002)
\bibitem{Mike} Tobar M.E. et al., Phys. Lett. {\bf A300}, 33, (2002).
\bibitem{Krupka2} Krupka J., Derzakowski K., Abramowicz A., Tobar M.E., Geyer R.G., IEEE Trans. on MTT, {\bf 47}, 6, 752, (1999).
\bibitem{Muller2} M\"uller H. et al., Phys. Rev. {\bf D67}, 056006, (2003).
\bibitem{KL} Kostelecky and Lane, Phys. Rev. {\bf D60}, 116010, (1999).
\bibitem{Bize} Bize S. et al., Proc. 6th Symp. on Freq. Standards and Metrology, World Scientific, (2002).
\bibitem{lss} Schmidt L.S., Metrologia {\bf 40}, in press, (2003).
\bibitem{Draper} Draper N.R. and Smith H., {\it Applied Regression Analysis}, Wiley, (1966).
\bibitem{ACES} Salomon C., et al., C.R. Acad. Sci. Paris, {\bf 2}, 4, 1313, (2001).
\bibitem{Lammerzahl2001} L\"ammerzahl C. et al., Class. Quant. Grav., {\bf 18}, 2499, (2001).
\end{thebibliography}
\end{document}